\def \yskip{\penalty-50\vskip3pt plus 3pt minus 2pt}
\def \reference{\par \yskip \noindent \hangindent .4in \hangafter 1}
\def \abc#1#2#3#4 {\reference#1, {\sl#2}, {\bf#3}, #4}
\def \blank {\lower 5pt\hbox to 0.75in{\hrulefill}}

\def \cm{~\rm{cm}}
\def \s{~\rm{s}}
\def \km{~\rm{km}}

\def \g{~\rm{g}}
\def \AU{~\rm{AU}}
\def \yrs{~\rm{yrs}}
\def \yr{~\rm{yr}}
\def \K{~\rm{K}}

\def \lae{\mathrel{<\kern-1.0em\lower0.9ex\hbox{$\sim$}}}
\def \gae{\mathrel{>\kern-1.0em\lower0.9ex\hbox{$\sim$}}}

\documentstyle[11pt,aasms,tighten,flushrt]{article}

\begin{document}
\small

\setcounter{page}{1}
\begin{center} \bf 
BACKFLOW IN POST-ASYMPTOTIC GIANT BRANCH STARS
\end{center}

\begin{center}
Noam Soker\\
Department of Physics, University of Haifa at Oranim\\
Oranim, Tivon 36006, ISRAEL \\
soker@physics.technion.ac.il 
\end{center}

\bigskip

\begin{center}
\bf ABSTRACT
\end{center}

 We derive the conditions for a backflow toward the central star(s)
of circumstellar material to occur during the post-asymptotic giant
branch (AGB) phase.
 The backflowing material may be accreted by the post-AGB star and/or
its companion, if such exists.
 Such a backflow may play a significant role in shaping the descendant
planetary nebula, by, among other things, slowing down the post-AGB
evolution, and by forming an accretion disk which may blow two jets.
 We consider three forces acting on a slowly moving mass element:
the gravity of the central system, radiation pressure, and fast wind
ram pressure.
 We find that for a significant backflow to occur, a slow dense
flow should exsist, such that the relation between the total mass
in the slow flow, $M_i$, and the solid angle it covers
$\Omega$, is given by  $(M_i/\beta) \gtrsim 0.1 M_\odot$,
where $\beta \equiv \Omega / 4 \pi$.
 The requirement for both high mass loss rate per unit solid angle
and a very slow wind, such that it can be decelerated and flow back,
probably requires close binary interaction.

\noindent
{\it Key words:}         
 Planetary nebulae: general
$-$ stars: AGB and post-AGB
$-$ stars: mass loss
$-$ circumstellar matter


\section{INTRODUCTION}

 More and more supporting observations (Sahai \& Trauger 1998;
Kwok, Su, \& Hrivnak 1998; Hrivnak, Kwok, \& Su 1999;
Kwok, Hrivnak, \& Su 2000; Huggins {\it et al.} 2001)
and theoretical considerations (Soker 1990; Soker 2001)
are accumulated in support of the view that significant shaping
of the circumstellar material takes place
just before, after, and during the transition from the asymptotic
giant branch (AGB) to the planetary nebula (PN) phase.
 Both the wind and radiation properties are significantly changed
during these stages.
 As the star is about to leave the AGB the mass loss rate increases
substantially, up to $\sim 10^{-5}-10^{-4} M_\odot \yr^{-1}$.
This wind was termed superwind by Renzini (1981);
because of confusion with other winds termed superwinds, hereafter
we'll refer to this wind as FIW, for {\it final intensive wind}.
After the star leaves the AGB the mass loss rate decreases
down to $\sim 10^{-8} M_\odot \yr^{-1}$, and its velocity
increases from $\sim 10 \km \s^{-1}$ to
few$\times 10^3 \km \s^{-1}$ at the PN phase.
 Simultaneously, the effective temperature increases, and the
post-AGB star starts to ionize the nebula around it, when by
definition the PN phase starts.
 The changes in the mass loss rate and velocity of the wind are
accompanied by a change in the wind geometry, e.g., jets and
bipolar structures are formed (Soker 1990; Sahai \& Trauger 1998;
Kwok {\it et al.} 2000).
 The change in the wind geometry may result from an intrinsic
processes in the AGB and post-AGB mass-losing star, or from
processes in an accreting companion (e.g., Soker 2001). 
 The increase in the wind velocity leads to collision of winds
(Kwok, Purton, \& Fitzgerald 1978), which can lead to instabilities
during the post-AGB (proto-PN) stage (Dwarkadas \& Balick 1998).
 As the central star starts ionizing the nebula, an ionization front
propagates outward, and plays a significant role in shaping the
nebula, both in the radial direction (e.g., Mellema \& Frank 1995;
Chevalier 1997; Sch\"onberner \& Steffen 2000), and in the
transverse directions (e.g., Mellema 1995; Soker 2000b).

 In the present paper we examine yet another process which may
occur during the transition from the AGB to the PN stage, i.e.,
after the FIW ceases and before ionization starts.
 This is a backflow of a fraction of the dense wind toward the
central star(s).
 This backflowing material may be accreted by the central star
and/or its companion.
 The idea of accreting backflowing material during the post-AGB
phase was raised before to explain and account for:
 a possible mechanism for the formation of jets from an
accretion disk (Bujarrabal, Alcolea, \& Neri 1998); 
a slower evolution along the post-AGB track
(Zijlstra {\it et a;.} 2000); 
and post-AGB stars depleted of refractory elements which
compose the dust particles (e.g., Van Winckel {\it et al.} 1998)
by accretion of a dust-depleted circumstellar gas 
(Waters, Trams \& Waelkens 1992), most likely in binary systems
(Van Winckel 1999). 

 The goal of the present paper is to explore the conditions
required for a backflow to occur such that it plays a
non-negligible role in the post-AGB evolution.
 The conditions are derived in $\S 2$, while the implications for
the processes mentioned above, as well as other processes,
together with a short summary, are in $\S 3$. 

\section{CONDITIONS FOR A BACKFLOW}

In this section we derive the conditions for a backflow to occur
during the post-AGB phase. 
 We assume that a very slow flow exists along some directions,
e.g., in the equatorial plane, and consider the conditions 
for some of this material to flow inward and be accreted by the
central system, before ionization starts.
 After being ionized, any dense cool gas will expand and will be
pushed outward by radiation and ram pressure (see below). 
 We do not consider the deceleration of the slowly outward 
moving mass element, but simply assume that if there is a 
slowly outward moving mass element, it will reach zero radial 
velocity at some radius. 

 We therefore consider a mass-element $M_i$ with a constant 
density $\rho_i$ within a solid angle $\Omega$ and a radial 
extention $\Delta r$ at a distance $r \gg \Delta r$ from the 
central star, such that 
\begin{equation}
M_i= \rho_i \Omega r^2 \Delta r.
\end{equation}
 The mass element can also be a shell where $\Omega=4 \pi$.
 If the sound crossing time $\Delta r /c_s$, where $c_s$ is the 
sound speed, is shorter than any other time scale in the process, 
we can take the shell to move more or less coherently.
 Three relevant forces are acting on the mass element in the
radial direction.
  The gravitational force of the central star(s)
\begin{equation}
f_g=-\frac {G M M_i}{r^2},
\end{equation}
where $M$ is the total mass of the central system, a binary system
or a single star.
 The fast wind blown by the central star and its radiation push 
outward.
 The force due to the fast wind's ram pressure, assuming it is
much faster than the slow wind velocity, is given by
\begin{equation}
f_w=\rho_w(r) v_w^2 \Omega r^2 =
\dot M_w v_w \beta,
\end{equation}
where $\rho_w(r)=\dot M_w/(4 \pi r^2 v_w)$, $v_w$ and $\dot M_w$ are
the density, velocity and mass loss rate (defined positively) of the 
fast wind, and $\beta \equiv \Omega/4 \pi$.
 The radiation imparts a force of
\begin{equation}
f_r=\frac {L_\ast}{c} \beta (1-e^{-\chi}),
\end{equation}
where $L_\ast$ is the luminosity of the central system, $c$ the
speed of light, and
\begin{equation}
\chi= \rho_i \kappa \Delta r = 7
\left(\frac {M_i}{0.01 M_\odot} \right)
\left(\frac {\kappa}{10 \cm^2 \g^{-1}} \right)
\left(\frac {r}{100 \AU} \right)^{-2} \beta^{-1}
\end{equation}
is the optical depth of the mass element, and we used equation (1)
for the density.
 We scale the opacity $\kappa(r)$ with a typical value for AGB stars
(Jura 1986; Winters {\it et al.} 2000), and
assume that the fast wind inner to the mass element
absorbs a negligible fraction of the radiation.
 For convenience we define three dimensionless variables.
The ratio of maximum radiation pressure to the fast wind ram pressure
\begin{equation}
q \equiv \frac{L_\ast/c}{\dot M_w v_w} =  
\left(\frac {L_\ast}{5000 L_\odot} \right)
\left( \frac{\dot M_w}{10^{-6} M_\odot \yr^{-1}} \right)^{-1}
\left( \frac{v_w}{100 \km \s^{-1}} \right)^{-1},
\end{equation}
which does not depend on $r$.
 We scale the fast wind mass loss and velocity as appropriate
for a post-AGB star before it starts ionizing the nebula.
 The ratio of the force due to radiation and wind to that of 
gravity depends on $r$, both through the dependence of 
optical depth $\chi$ and gravity on $r$. 
 We define $\eta$ to be this ratio at a scaling radius $r_0$
\begin{equation}
\eta= \frac {\dot M_w v_w}{G M M_i} k(r_0) r_0^2, 
\end{equation}
where
\begin{equation}
k(r_0) \equiv \beta [1+q(1-e^{-\chi})].
\end{equation}
 
The equation of motion for the mass element can be written as
\begin{equation}
\frac {d^2r}{dt^2}= \frac{G M}{r_0^2} 
\left( \eta -\frac{r_0^2}{r^2} \right).
\end{equation}
We take $r_0$ to be the radius at which the radial velocity of the
mass element is zero.
When there is only gravity, the free-fall time from $r=r_0$ to
the center $r=0$, with $v(r_0)=0$, is
\begin{equation}
t_{ff}=\frac{\pi}{2^{3/2}} \frac{r_0^{3/2}}{(G M)^{1/2}}= 1400
\left( \frac{r_0}{400 \AU} \right)^{3/2} 
\left( \frac{M}{1 M_\odot} \right)^{-1/2} \yr.
\end{equation}
 We define the dimensionless variables
\begin{equation}
\tau=t/t_{ff} \qquad {\rm and} \qquad x=r/r_0,
\end{equation}
 and write the equation of motion (9) in the form
\begin{equation}
\frac {d^2x}{d\tau^2}= \frac {\pi^2}{8}(\eta-x^{-2}).
\end{equation}
 Assuming that $\eta$ is constant and does not depend on $x$
allows us to integrate once the last equation to give the velocity
\begin{equation}
v \equiv \frac {dx}{d\tau}= - \frac {\pi}{2}
(\eta x+x^{-1}-1-\eta)^{1/2},
\end{equation}
where we substituted the initial condition $v(1)=0$.
 This can be integrated analytically for constant values of $\eta=1$
and $\eta=0$.
  The case $\eta=0$ gives the free fall solution.
 The time left for the object to fall from $x$ to $x=0$ is given by
\begin{equation}
\tau_f(x)= \frac{2}{\pi} \left(\sin^{-1}x^{1/2}-[x(1-x)]^{1/2}\right).
\end{equation}
 As expected from our scaling $\tau(1)=1$, i.e., the free fall time
from $r=r_0$ to $r=0$ is $t_{ff}$. 
 For $\eta=1$ and with $v(1)=0$, the fall time from $x=1$
is infinite.
This is because the wind and radiation outward-acceleration equals
the gravity inward-acceleration.
 However, the time left to fall from $x$ to $x=0$, with $v(1)=0$, 
is finite, and is given by 
\begin{equation}
\tau_f(x)= \frac{2}{\pi} \left( \ln [(1+x^{1/2})/(1-x^{1/2})^{-1}] 
- 2 x^{1/2} \right).
\end{equation}
 
Comparing the last two equations, we find as expected, that
radiation pressure and wind's ram pressure slow down the
inflow. 
For example, for $\eta=0$, i.e., no wind and radiation pressure,
the time left to fall from $x=0.9$ to $x=0$ is $\tau=0.604$,
while for $\eta=1$ it is $\tau= 1.11$.
In both cases the initial condition is $v(1)=0$.
 These cases are less interesting, since for $\eta=1$ the flow 
actually stagnates at $x=1$. 
 More interesting is the case of $v(1)=0$ and $0< \eta <1$,
since for $\eta>1$ it will be push away from the center.
 We numerically integrated equation (13) for these conditions
and for constant values of $\eta$.
We find the backflow time to be $\tau_f(1)=1.28$, $1.52$, and
$2.08$, for $\eta=0.5$, $0.7$, and $0.9$, respectively.
 For $\eta=0.8$, for example, the fall back times from
$x_i=1$, $0.9$ and $0.8$ are $\tau_f=1.72$, $1.23$ and $0.925$, 
respectively,  where $v(x_i)=0$ in these cases. 
For $\eta=0$ the backflow times for the same initial conditions
are $1$, $0.85$, and $0.72$, respectively.
 The conclusion from the numerical values cited above is that
the typical backflow time from $r \simeq r_0$ is the free fall time 
$\sim t_{ff}$ at $r_0$, but because of the radiation and wind
pressures the region from which this is the backflow time is much
larger than the $\eta=0$ cases, extending from $r_0$ down to
$\sim 0.7-0.8 r_0$ for $\eta \gtrsim 0.8$.
 So the question is what is the value of $r_0$ for which $\eta=1$.
Below and close to this radius the gas falls back in a 
time $\sim t_{ff}$, while it is accelerated away for larger radii.
From equation (7) we find
\begin{equation}
r_0 =
\left( \frac {\eta GMM_i}{k \dot M_w v_w } \right)^{1/2}
= 430 \AU  \left[
\left( \frac{M} {1 M_\odot} \right)
\left( \frac{M_i} {0.01M_\odot} \right)
\left( \frac{\dot M_w} {10^{-6} M_\odot} \right)^{-1}
\left( \frac{{\it v_w}} {100 \km \s^{-1}} \right)^{-1}
\left( \frac{k}{0.1} \right)^{-1} \eta \right]^{1/2}
\end{equation}

 For the backflowing mass to influence the evolution significantly,
we required the fall back time to be $\gtrsim 1000 \yr$. 
 For the typical parameters used in equations (5) and (6) we find
from equation (8) that $k(r) \sim 2 \beta$; using the time scale 
given by equation (10) in equation (16) gives the desired condition 
\begin{equation}
\frac{M_i} {\beta } \gtrsim 0.1 M_\odot 
\end{equation}
where as before, $\beta\equiv \Omega /4 \pi$, and 
$\Omega$ is the solid angle covered by the dense 
backflowing material.
 The last condition is limited by a maximum density, since a large 
value of the backflowing mass $M_i$ and small value for $\beta$ means
a very high density. 
 We now estimate a reasonable value for the density.
We assume a very slow equatorial flow, with a speed of
$v_s \sim 1 \km \s^{-1}$ and with a mass loss rate per unit solid angle 
of $\dot m_s = \dot M_s / 4 \pi$.
The density of the mass elements formed by this wind is 
\begin{equation}
\rho_{iw} = \frac {\dot m_s}{r^2 v_s} = 1.4 \times 10^{-15}
\left( \frac{ \dot M_s} {10^{-3} M_\odot \yr^{-1}} \right) 
\left( \frac{ r } {400 \AU} \right) ^{-2}
\left( \frac{ v_s } {1 \km \s^{-1}} \right) ^{-1} \g \cm^{-3}.
\end{equation}
 The minimum density is that for which the fast wind compresses the
dense cool wind such that the ram pressure $\rho_w v_w^2$ equals
the thermal pressure of the cool gas. 
 For a molecular gas, we find this density to be
\begin{equation}
\rho_{ip} = \rho_w \frac{v_w^2}{c_i^2} = 10^{-16}
\left( \frac{\dot M_w} {10^{-6} M_\odot} \right)
\left( \frac{v_w} {100 \km \s^{-1}} \right)
\left( \frac{ r } {400 \AU} \right) ^{-2}
\left( \frac{ T_i } {300 \K} \right) ^{-1} \g \cm^{-3}.
\end{equation}
where $c_i$ and $T_i$ are the isothermal sound speed and temperature,
respectively, of the cool gas. 
 We find that a density of $\rho_i \sim 10^{-15} \g \cm^{-3}$ is 
reasonable. 
 Using equation (1) and the definition of $\beta$ gives 
\begin{equation}
\frac{M_i} {\beta } = 0.14
\left( \frac {\rho_i}{10^{-15} \g \cm^{-3}} \right)
\left( \frac {r}{400 \AU} \right)^3
\left( \frac {\Delta r}{0.1 r } \right) M_\odot.
\end{equation}
 Condition (17) is met for the density given by equation (18),
but this requires a very high mass loss rate per unit solid angle.
If $\beta=0.1$ this requires a total mass loss rate of
$\dot M=10^{-4} M_\odot \yr^{-1}$, but concentrated 
in particular directions, probably in the equatorial plane. 
 All these considerations strongly suggest an equatorial dense
and slow flow, such as expected in a close binary system 
(Mastrodemos \& Morris 1999; Soker 2000a).
 A very fast rotation can also form such a wind (Bjorkman \& Cassinelli
1993), but then a binary companion is needed to substantially
spin-up the envelope.

\section{IMPLICATIONS AND SUMMARY}

 Despite the assumptions and simplifications in deriving 
condition (17), we feel that the results obtained in the
previous section and the implications discussed in this section
are quite general.
 We find that for a backflow to occur on a time scale of
$t_{\rm acc} \gtrsim 10^3 \yr$ after the termination of the AGB,
so that it has a non-negligible role in the post-AGB evolution,
the following conditions should be met:
(1) The total backflowing mass should be larger than the
combined mass lost in the wind and that burned in the core.
For a post-AGB mass loss rate of $\sim 10^{-6} M_\odot \yr^{-1}$
the nuclear burning is negligible, and the total required mass
is $M_{\rm acc} \simeq 10^{-3} (t_{\rm acc} / 1,000 \yr) M_\odot$.
(2) The backflowing mass should have a very low, 
$\sim 1 \km \s^{-1}$, terminal velocity, so that eventually it will
be decelerated to zero velocity, and flow back.
(3) Condition (17) on the ratio of the mass of the 
mass-element and the solid angle it covers
$\beta= \Omega / 4 \pi$ should be met.  
(4) For reasonable densities (eq. 18), and the required mass,
we find (eq. 20) that the mass range is 
$M_i \simeq 0.1-10^{-3} M_\odot$, and the appropriate solid 
angle covered by the backflowing mass is 
$1 < \beta \lesssim 10^{-2}$.
 We can take the typical values to be $M_i \simeq 0.01 M_\odot$
and $\beta=0.1$.

 These mass loss properties required that 
($i$) the flow be concentrated along particular directions,
and ($ii$) have an inefficient acceleration by the stellar radiation.
 In a previous paper (Soker 2000a) two mechanisms which lead
to such a flow were discussed.
 In the first mechanism proposed by Soker (2000a) magnetic cool spots
(as in the Sun) are formed on the surface of slowly rotating AGB stars.
 The lower temperature above the spots enhances dust formation,
which during the final intensive wind (FIW) may lead to an
optically thick wind.
 This means an inefficient radiative acceleration, hence a slow flow
above the spot. 
If the spot is small, material from the surrounding flows into the shaded
region and accelerates the slow flow.
 However, if the spot is large, material from the surroundings will
not accelerate the flow much, and it will stay slow.
 In the present paper we find that the condition for the slowly
moving material to flow back is that the spot has
$\beta \gtrsim 0.01$, which for a circular spot means a radius
of $R_{\rm spot} \gtrsim 0.2 R_\ast$, where $R_\ast$ is the stellar radius.
 The required mass in the slow flow is given by equation (17).

 In the second mechanism a high density in the equatorial plane is
formed by a binary interaction, where the secondary star is close to, but
outside the AGB envelope.
This mechanism is supported by observations, e.g., 
slowly moving equatorial gas is found around several binary post-AGB
stars (Van Winckel 1999, and references therein). 
 In the process proposed by Soker (2000a) the strong interaction
between the two stars forms a dense equatorial outflow which is
optically thick, leading to an inefficient radiative acceleration
and a very slow equatorial flow. 
 For a very massive and significant backflow, with a mass
of $M_i \gtrsim 0.01 M_\odot$ and $\beta \gtrsim 0.1$, to occur,
a binary mechanism is required.
 For the cool spots model to form such a massive slow flow,
several large spots are required.
 This means a strong magnetic activity, which probably requires
the AGB star to be spun-up by a stellar companion. 
 We therefore argue that in both mechanisms a binary companion is 
required to cause a massive flow, such that it may last for 
$t_{\rm acc} \gtrsim 10^3 \yrs$, possibly by as long as $\sim 10^4 \yrs$ 
in extreme cases. 

 Such a backflow may have the following effects.
 If it has a large specific angular momentum, as expected in
strongly interacting binary systems, the backflowing material may form
an accretion disk around one or two of the two stars.
 The disk(s) may blow jets or collimated fast winds (CFWs),
 which will play a significant role in shaping the circumstellar material.
 Such a possibility was briefly mentioned by
Bujarrabal {\it et al.} (1998) for the proto-PN M1-92.
 The accreted mass may slow down the post-AGB evolution, as
suggested by Zijlstra {\it et al.} (2000) for some OH/IR stars.
 Zijlstra {\it et al.} (2000) termed these stars retarded stars,
and argue for a delay as long as $10^4 \yrs$ by accretion from
a near-stationary reservoir.
They bring supporting observations, but don't consider the formation
of such a reservoir of mass.
 The results of the present paper put the idea of 
Zijlstra {\it et al.} (2000) on a more solid ground.
 Another effect attributed to backflow accretion
is the formation of post-AGB stars depleted of refractory elements
which compose the dust particles (e.g., Van Winckel {\it et al.} 1998;
Waters, Trams \& Waelkens 1992; Van Winckel 1999).
 The separation of dust from the gas was not considered in the
present paper.

 Finally, we speculate on another plausible effect of the dense 
backflowing gas.
 The central star's wind is shocked when it hits the dense material.
 If some dense backflowing blobs survive long into the PN phase, when
the central star's wind velocity is $\gtrsim 10^3 \km \s^{-1}$, then there
will be a hard X-ray emission from the post-shock fast wind material.
 The dense blob will be close to the central star, making the
hard X-ray emitting region hard to resolve.
  It is not clear if this compact hard X-ray emitting region can
explain the recent {\it Chandra} observations of a ``point source''
in the centers of the Helix (NGC 7293) and Cat's Eye (NGC 6543) PNe
(Guerrero {\it et al.} 2000).

{\bf ACKNOWLEDGMENTS:} 
I thank Benzion Kon for his help with the mathematics.
  This research was supported by a grant from the US-Israel
Binational Science Foundation.

\end{document}